\documentclass[prb,twocolumn,showpacs,aps,superscriptaddress,floatfix]{revtex4}
\usepackage{amsmath}
\usepackage{amssymb}
\usepackage{amsfonts}
\usepackage{bm}
\usepackage{graphicx}
\usepackage{color}
\usepackage{hyperref}
\begin{document}

\title{Scattering on a rectangular potential barrier in nodal-line Weyl
semimetals}

\author{ D.A. Khokhlov}
\affiliation{Moscow Institute for Physics and Technology (State
University), Moscow region, 141700 Russia}
\affiliation{Dukhov Research Institute of Automatics, Moscow, 127055
Russia}

\author{A.L. Rakhmanov}
\affiliation{Moscow Institute for Physics and Technology (State
University), Moscow region, 141700 Russia}
\affiliation{Dukhov Research Institute of Automatics, Moscow, 127055
Russia}
\affiliation{Institute for Theoretical and Applied Electrodynamics, Russian
Academy of Sciences, Moscow, 125412 Russia}

\author{A.V. Rozhkov}
\affiliation{Moscow Institute for Physics and Technology (State
University), Moscow region, 141700 Russia}
\affiliation{Institute for Theoretical and Applied Electrodynamics, Russian
Academy of Sciences, Moscow, 125412 Russia}

\begin{abstract}
We investigate single-particle ballistic scattering on a rectangular
barrier in the nodal-line Weyl semimetals. Since the system under study has
a crystallographic anisotropy, the scattering properties are dependent on
mutual orientation of the crystalline axis and the barrier. To account for
the anisotropy, we examine two different barrier orientations. It is
demonstrated that, for certain angles of incidence, the incoming particle
passes through the barrier with probability of unity. This is a
manifestation of the Klein tunneling, a familiar phenomenon in the context
of graphene and semimetals with Weyl points. However, the Klein tunneling
in the Weyl-ring systems is observed when the angle of incidence differs
from 90$^\circ$,
unlike the cases of graphene and Weyl-point semimetals.
The reflectionless transmission also occurs for the so-called `magic
angles'. The values of `the magic angles' are determined by geometrical
resonances between the barrier width and the de~Broglie length of the
scattered particle. In addition, we show that under certain conditions the
wave function of the transmitted and reflected particles may be a
superposition of two plane waves with unequal momenta.  Such a feature is a
consequence of the non-trivial structure of the iso-energy surfaces of the
nodal-line semimetals.
\end{abstract}
\date{\today}

\maketitle

\section{Introduction}

Topological semimetals (TSMs) were predicted
theoretically~\cite{Burkov2016,PhysRevB.84.235126,PhysRevB.92.081201,
PhysRevB.93.085138,PhysRevLett.107.127205,PhysRevLett.115.036806,Weng2015}
and then discovered
experimentally~\cite{neupane2014observation,xie2015new,PhysRevB.93.201104,
PhysRevLett.115.036807}. 
The main feature of the TSMs is that the valence and conduction bands
intersect in several (nodal) points or closed (nodal) lines in momentum
space~\cite{PhysRevB.84.235126,PhysRevLett.107.127205,PhysRevB.83.205101}.
Thus, in contrast to a usual metal with a two-dimensional Fermi surface,
the Fermi surface of a three-dimensional TSM is reduced to a finite set of
points or curves. Specifically, for the nodal-line semimetals, which we
discuss below, the Fermi surface is shrunk to a curve.

Within a simplest theoretical
framework~\cite{PhysRevB.93.085138},
the nodal line (or ring) of a Weyl semimetal is a closed plane curve in a
three-dimensional Brillouin
zone~\cite{PhysRevB.84.235126,PhysRevB.92.081201,PhysRevB.93.085138,Fang2016}.
In the plane of the nodal line, which we will also refer to as the basal
plane, the kinetic energies of the electrons and holes are proportional to
the square of the distance from the nodal line. As the momentum deviates
from the nodal line in the direction normal to the basal plane, the energy
variation is proportional to this deviation.

Since the spectrum of the nodal-ring semimetals is rather peculiar, one can
expect some unusual scattering effects in these materials. For example, the
electron and hole bands touch each other near the Fermi energy in the TSMs,
and even weak spatial variation of the potential energy can lead to the
interband transitions. Consequently, a single-particle scattering problem
must take into account both electron and hole states. This is a necessary
condition for observation of the Klein phenomenon, that is, a process in
which a particle passes through high and long potential barrier without
reflection~\cite{klein1929reflexion,klein_tunn2006}.
Klein tunneling was described
theoretically~\cite{klein_tunn2006,rozhkov2016electronic}
and observed
experimentally~\cite{young_fabry_perot}
in graphene. It suppresses the backscattering, thus contributing to the
increase of the graphene electron mean free path.
It is known that the Klein phenomenon requires linear single-particle
spectrum. Since electron dispersion in the nodal-ring semimetals is linear
in some directions in
$\mathbf{k}$-space,
the Klein tunneling might be expected in these materials under certain
conditions.

Here we study the scattering of an electron by a step-wise potential
barrier in the ballistic regime. Four different cases are considered:
barriers with finite and infinite width, both perpendicular and parallel to
the basal plane. We demonstrate that reflectionless tunneling is possible
for both orientations of the barrier. Similar to the
graphene~\cite{klein_tunn2006},
the perfect transmission in the nodal-ring semimetals is associated with
both Klein tunneling and `magic angles'
resonances~\cite{chiral_tunn_tudorovskiy2012}.
The Klein phenomenon in the nodal-ring semimetals is observed if the
incident angle of the particle differs from
90$^\circ$.
This is dissimilar to the case of graphene and Weyl-point semimetals, where
this effect exists only for normal incidence.

Another interesting feature of the scattering in the studied materials is
the emergence of two transmission and two reflection channels for a single
incident plane wave. To characterize such a scattering one needs two
transmission and two reflection coefficients. The transmitted (reflected)
particles in different transmission (reflection) channels have different
momenta. This property is in stark contrast with the scattering of free
non-relativistic electrons, where the momentum of the outgoing particles
are uniquely fixed by the momentum of the incident particle. The existence
of the multiple scattering channels is a consequence of the complicated
dispersion structure of a nodal-ring semimetal.

The paper is organized as follows. In
Sec.~\ref{sec::model}
we briefly discuss the theoretical model of a nodal-ring semimetal. This
model is used in
Sec.~\ref{sec::parall}
to study the scattering on a barrier parallel to the basal plane. The
barrier perpendicular to the basal plane is discussed in
Sec.~\ref{sec::perp}. 
Summary and conclusions are in
Sec.~\ref{sec::discussion}.

\section{Model} \label{sec::model}

We write the Hamiltonian of the system in the following
form~\cite{PhysRevB.93.085138}:
\begin{equation}\label{eq::hamilt}
\hat{H}(\mathbf{k})=(m-Bk_{\bot}^2)\sigma_x+k_z \sigma_z + U\sigma_0,
\end{equation}
where
$\mathbf{k}=(k_x,k_y,k_z)$
is the single-particle momentum, and scalar
$k_{\bot}$
is determined by the formula
$k_{\bot}^2=k_x^2+k_y^2$,
coefficient
$m$
is an analog of the rest mass, the quantity
${1}/(2B)$
is
an inertial mass for the motion in the
$xy$-plane, and
$U$
is the potential
energy. Matrix
$\sigma_0$
is the 2x2 unity matrix and
$\sigma_{x,z}$
are
the Pauli matrices. We set
$\hbar$
and
$v_{F}$
in
$z$~direction equal to
one. The spectrum of Hamiltonian~(\ref{eq::hamilt}) is \begin{equation}
\varepsilon_{\mathbf{k}}^{\rm e,h}=U\pm \sqrt{(m-Bk_{\bot}^2)^2+k_z^2},
\label{2}
\end{equation}
where label `e' (`h') corresponds to electrons (holes). If the potential
energy is zero, the solutions of the equation
$\varepsilon_{\mathbf{k}}=0$
forms a circle of radius
$k_\bot=\sqrt{m/B}$
in the
$xy$-plane. This circle
is the nodal line of the model, and
$xy$-plane is the basal plane.
Normalized eigenfunctions of the Hamiltonian~(\ref{eq::hamilt}) are spinors
\begin{eqnarray}\label{3}
\psi_\mathbf{k}\! =\! C_{\mathbf{k}}\begin{pmatrix}1\\
\chi_{\mathbf{k}}\\
\end{pmatrix},\,\,
\chi_{\mathbf{k}}\!=\!\frac{\varepsilon\!-\!U\!-\!k_z}{m-Bk_{\bot}^2},
\,\,
C_{\mathbf{k}}\!=\!\frac{1}{\sqrt{1\!+\!\chi_{\mathbf{k}}^2}}.
\end{eqnarray}
Following a standard
procedure~\cite{landau1981quantum},
we can calculate the probability current associated with a plane wave
$\psi (\mathbf{r})=a\psi_{\bf k} e^{i\mathbf{kr}}$:
\begin{equation}\label{eq::current}
\mathbf{j}= |a|^2
\begin{pmatrix}		
	-4\chi_{\mathbf{k}} B k_x \\
	-4\chi_{\mathbf{k}} B k_y \\
	1-\chi^2_\mathbf{k} \\
\end{pmatrix}.
\end{equation}
This current is invariant with respect to rotation in the
$xy$-plane. We
will use
Eq.~(\ref{eq::current})
below to choose a correct structure of the outgoing waves and to define
properly transmission and reflection coefficients.

\section{Barrier parallel to basal plane}\label{sec::parall}

\begin{figure}
\center{\includegraphics[width=0.9\linewidth]{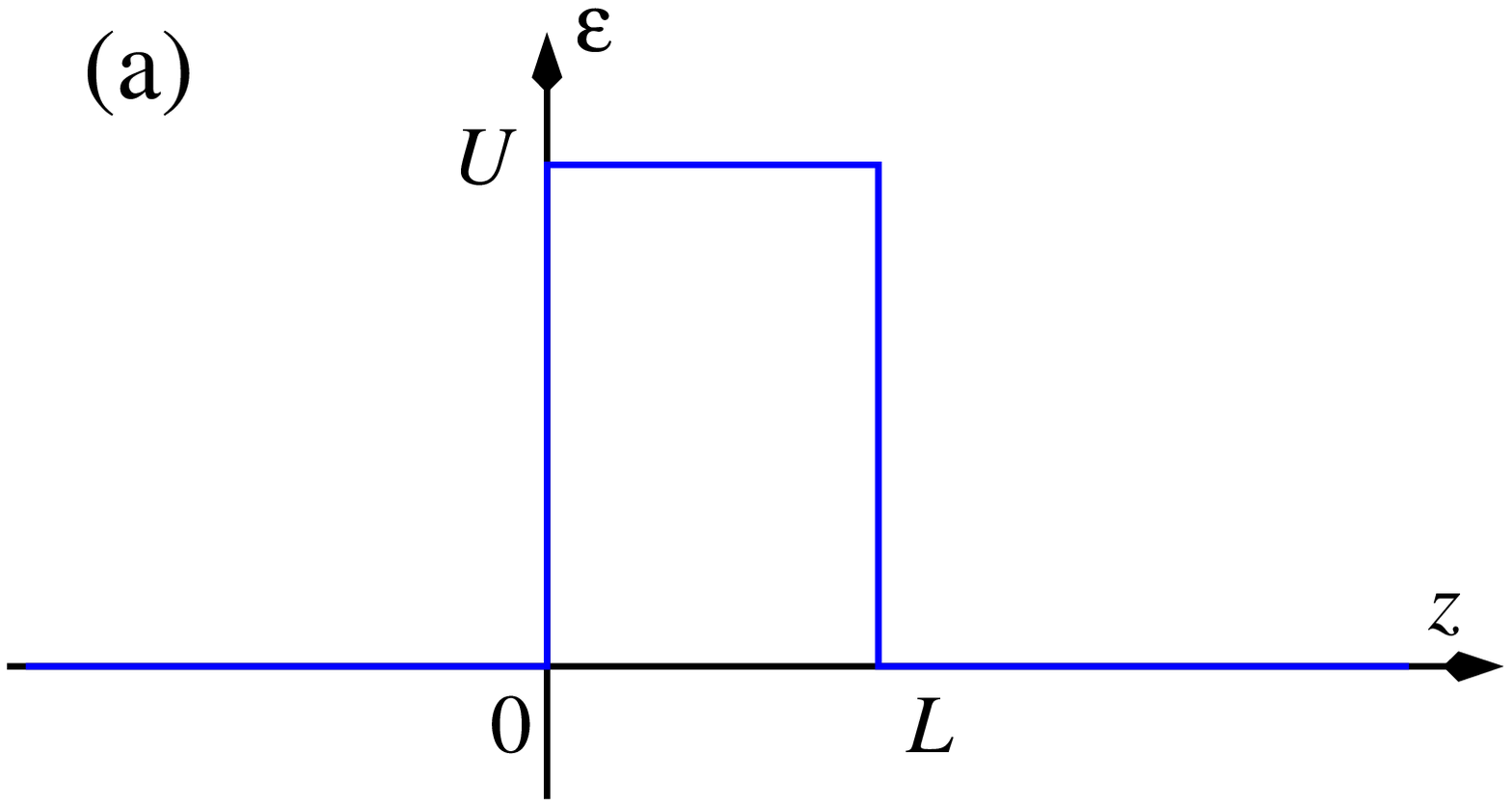}}
\center{\includegraphics[width=0.9\linewidth]{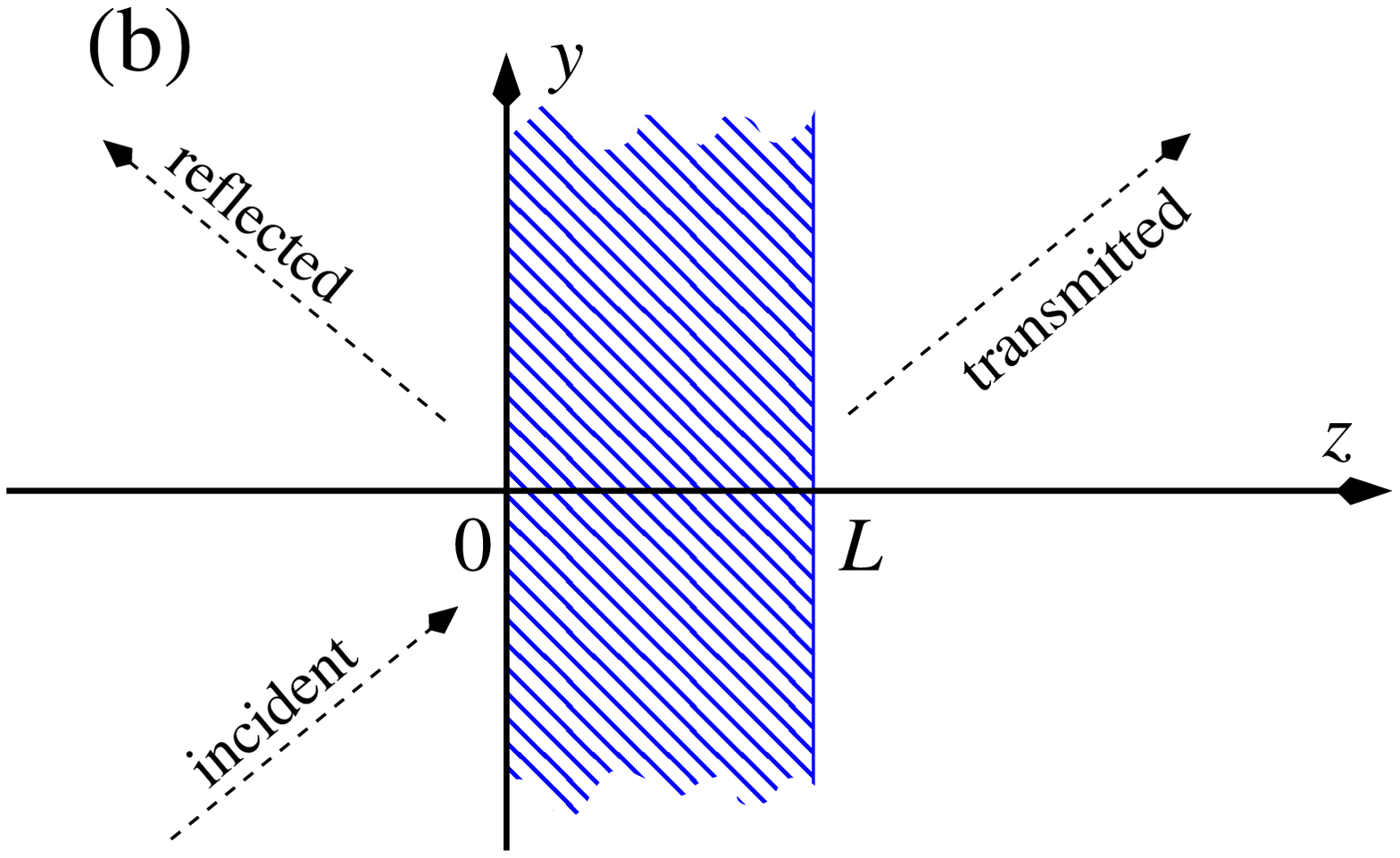}}
\center{\includegraphics[width=0.9\linewidth]{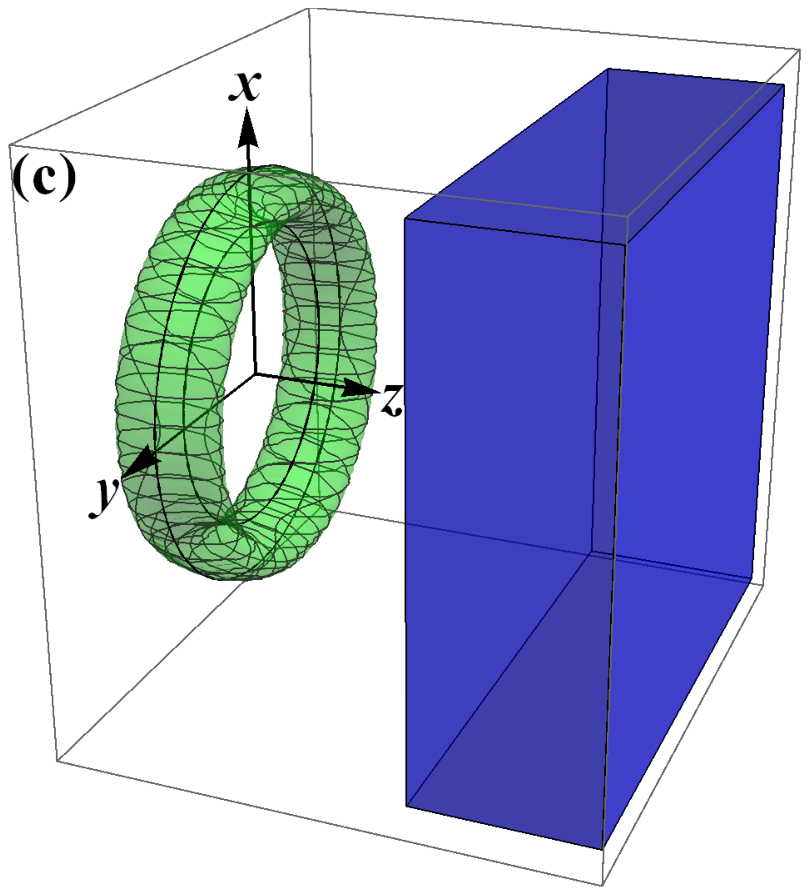}}
\caption{Barrier parallel to the basal plane. Panel~(a): Potential energy
$U(z)$.
It is finite and constant for
$0<z<L$.
Otherwise, it is zero.
Panel~(b): Incident, reflected, and transmitted waves near the barrier. The
wave vectors of the incoming and outgoing waves are shown schematically by
the dashed lines with arrows. The barrier is represented by the blue hatched
area.
Panel~(c): Relative orientations of the barrier and the iso-energy surface.
The iso-energy surface is defined by the equation
$\varepsilon^2 = {(m-Bk_{\bot}^2)^2+k_z^2}$.
It is shown by (green) torus in the reciprocal space. The barrier is shown
as blue parallelepiped. 
\label{Fig1}}
\end{figure}

\begin{figure}[t]
\center{\includegraphics[width=1\linewidth]{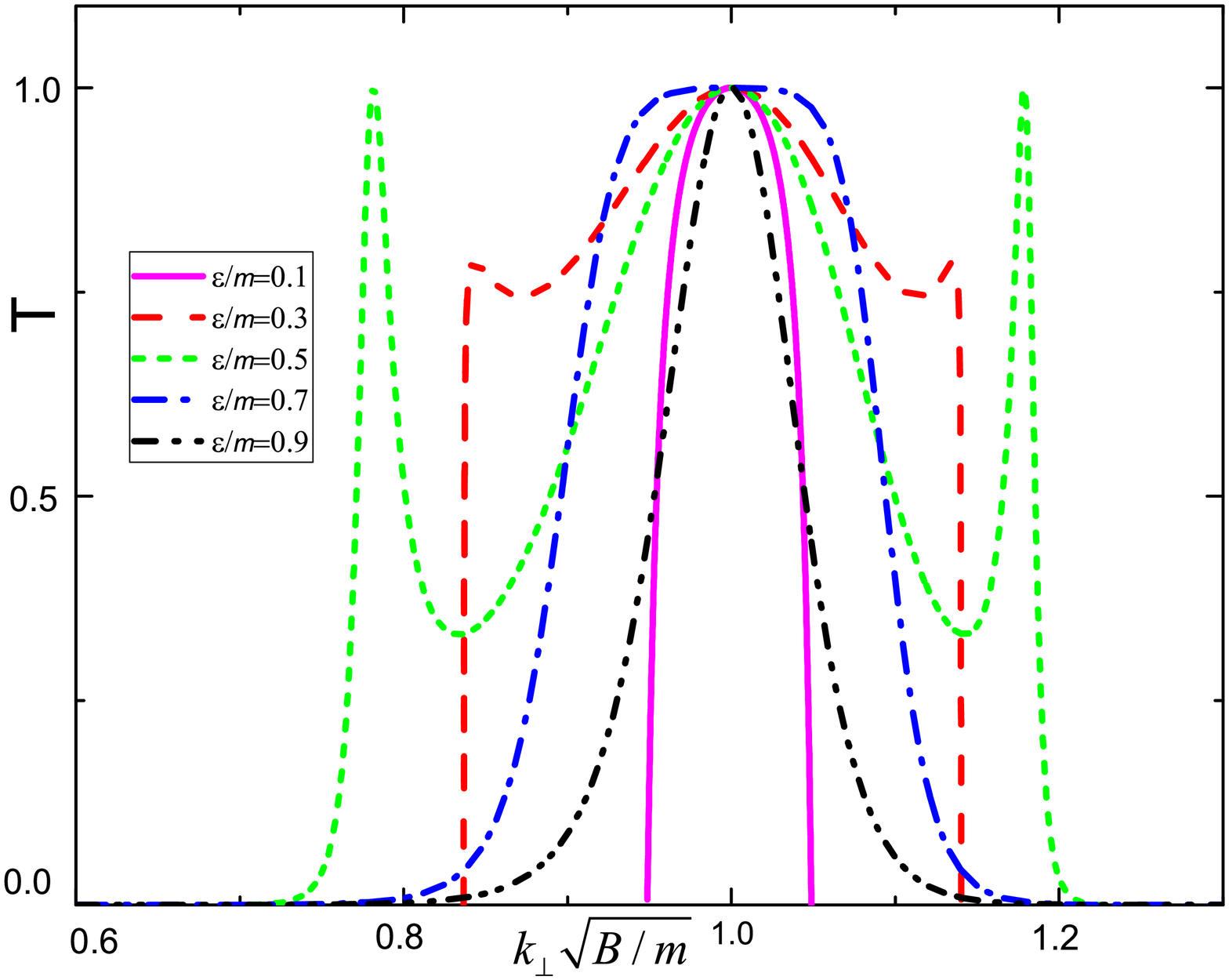}}
\caption{Transmission coefficient $T$ as a function of the dimensionless
transverse momentum
$k_{\bot}\sqrt{B/m}$
for different values of the ratio
$\varepsilon/m$
(see legend in the figure). The curves are calculated for
$U/m=1$,\,
$mL=10$,\, 
and
$Bm=1$.
Momentum
$k_{\bot}$
is limited by the conditions
$\sqrt{(m-\varepsilon)/B}<k_{\bot}<\sqrt{(m+\varepsilon)/B}$,
which guarantees that a propagating solution (${\rm Im}\, k_z=0$)
exists. When
$k_{\bot}\sqrt{B/m}=1$,
a perfect transmission due to the Klein tunneling is observed. In addition,
the reflectionless tunneling at the so-called `magic angles' is also
possible.
\label{fig::T_vs_k_perp}
}
\end{figure}

\begin{figure}[t]
\center{\includegraphics[width=1\linewidth]{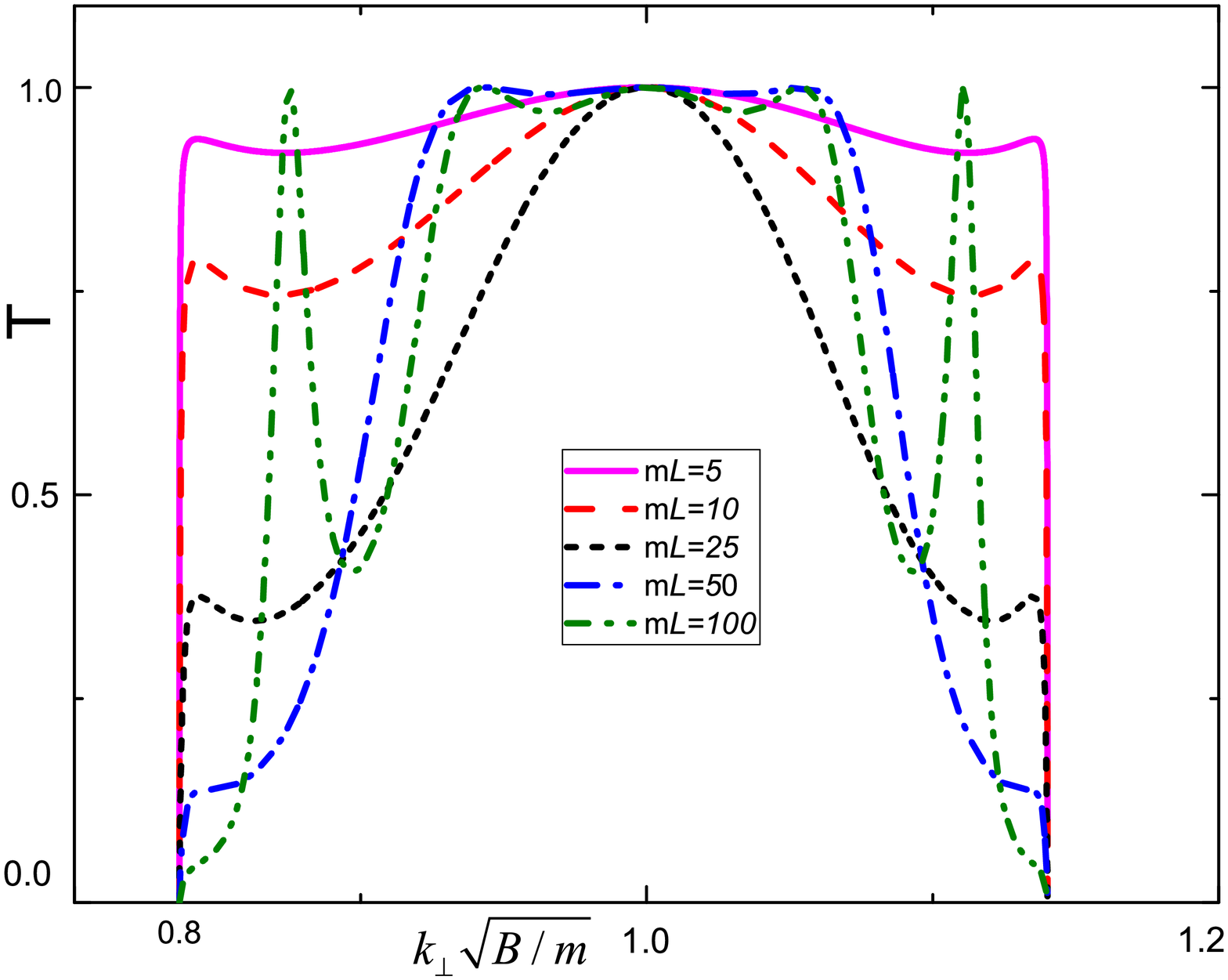}}
\caption{Transmission coefficient $T$ as a function of the dimensionless
momentum
$k_{\bot}\sqrt{B/m}$
for different dimensionless barrier widths $mL$ (see legend in the figure).
The curves are calculated for
$U/m=1$,
$\varepsilon/m=0.3$,
and
$Bm=1$.
When
$k_{\bot}\sqrt{B/m}=1$,
perfect
transmission due to the Klein tunneling is observed. In addition, the
reflectionless tunneling at the so-called `magic angles' is also possible.
The latter becomes more pronounced for wider barriers.
\label{fig::T_vs_L}
}
\end{figure}

First we consider the situation when the barrier is parallel to the basal
plane, see
Fig.~\ref{Fig1}. The potential energy
$U(z) = U [ \theta(z) - \theta(z-L)]$,
where 
$\theta(z)$
is the Heaviside step-function, depends on
$z$~coordinate only. Barrier width in the $z$ direction is equal to $L$.
In the $x$ and $y$ directions the barrier extends to infinity. Thus, the
momentum components
$k_x$
and
$k_y$
are conserved. Further, we assume that the incident particle is an electron
(not a hole). The barrier divides the space into three regions (to the left
of the barrier, to the right of the barrier, and under it). The wave
function in these three regions is
\begin{eqnarray}
\label{eq::match1}
\nonumber \psi &=& C_{+} e^{ik_zz}
\begin{pmatrix}
	1\\
	\chi_{+}\\
\end{pmatrix}
+ rC_{-}e^{-ik_zz}
\begin{pmatrix}
	1\\
	\chi_{-}\\
\end{pmatrix}\!\!,\, z<0,
\\
\psi &=& De^{iq_zz}\!
	\begin{pmatrix}
		1\\
		\phi_{+}\\
	\end{pmatrix}
	\!+\! Fe^{-iq_zz}\!
	\begin{pmatrix}
		1\\
		\phi_{-}\\
	\end{pmatrix}\!\!,\,\, 0<z<L,
\\
\nonumber \psi &=& t\,C_{+} e^{ik_z(z-L)}
	\begin{pmatrix}
		1\\
		\chi_{+}\\
	\end{pmatrix}\!\!,\,\, z>L.
\end{eqnarray}
Here
\begin{eqnarray}
\label{eq::kz_def}
k_z&=&\sqrt{\varepsilon^2-(m-Bk_{\bot}^2)^2},\\
\label{eq::qz_def}
q_z &=& \sqrt{(\varepsilon-U)^2-(m-Bk_{\bot}^2)^2}.
\end{eqnarray}
The quantities
$\pm k_z$
and
$\pm q_z$
are the
$z$-components of the wave vectors outside and inside the barrier,
respectively. Energy of the incident electron is 
$\varepsilon$,
and
\begin{eqnarray}
\chi_\pm &=& \chi( \pm k_z)=\frac{\varepsilon \mp k_z}{m-Bk_{\bot}^2},
\\
\nonumber
\phi_\pm &=&\phi(\pm q_z)=\frac{\varepsilon-U \mp q_z}{m-Bk_{\bot}^2},\\ \nonumber
\\
\nonumber
C_\pm &=& (1 + \chi_\pm^2)^{-1/2}.
\end{eqnarray}
Factor
$e^{ik_x x + i k_y y}$,
common for all three wave functions, is
omitted for brevity. To describe a particle propagating freely outside the
barrier, the wave function must have purely real
$k_z$,
or, equivalently,
\begin{equation}
{\frac{m-\varepsilon}{B}}<k^2_{\bot}<{\frac{m+\varepsilon}{B}}.
\label{eq::window_kperp}
\end{equation}
Transmission and reflection coefficients
$T=|t|^2$
and
$R=|r|^2$
obey the usual relation
$T+R=1$.
To derive $r$ and $t$ we should match $\psi$ at
$z=0$
and
$z=L$,
accounting for the continuity of the probability current. In this way we
derive
\begin{align}
		C_{+}
		\begin{pmatrix}
			1\\
			\chi_+\\
		\end{pmatrix}
		+ rC_-
		\begin{pmatrix}
			1\\
			\chi_-\\
		\end{pmatrix}
		=
		D
		\begin{pmatrix}
			1\\
			\phi_+\\
		\end{pmatrix}
		+
		F
		\begin{pmatrix}
			1\\
			\phi_-\\
		\end{pmatrix},\nonumber	\\
		tC_+
		\begin{pmatrix}
			1\\
			\chi_+\\
		\end{pmatrix}
		=
		De^{iq_zL}
		\begin{pmatrix}
			1\\
			\phi_+\\
		\end{pmatrix}
		+
		Fe^{-iq_zL}
		\begin{pmatrix}
			1\\
			\phi_-\\
		\end{pmatrix}.
        \label{10}
	\end{align}
Solving
system~(\ref{10})
one obtains the expression for $r$
\begin{equation}
\label{eq::refl1b}
r\!=\!\frac{(U-k_z)^2-q_z^2}{k_z^2+q_z^2-U^2 + 2 i k_z q_z {\rm cot} (q_zL)}
\sqrt{\frac{\varepsilon+k_z}{\varepsilon-k_z}}.
\end{equation}
The dependence of the transmission coefficient
$T=1-|r|^2$
on the transverse momentum
$k_\bot$
is calculated using
Eq.~\eqref{eq::refl1b}. 
The results for several energies $\varepsilon$ are shown in
Fig.~\ref{fig::T_vs_k_perp}.
The same dependence for different barrier widths $L$ is presented in
Fig.~\ref{fig::T_vs_L}.
We see that for certain parameter values the transmission is perfect:
$T=1$,
or, equivalently,
$r=0$.
Note that
$k_\bot$
varies in the interval defined by the 
conditions~\eqref{eq::window_kperp}.

As it follows from
Eqs.~\eqref{2},
\eqref{eq::kz_def},
\eqref{eq::qz_def}, 
and~(\ref{eq::refl1b}), 
the disappearance of the reflected wave 
($r=0$)
occurs when either of two different conditions is satisfied. The first one
is
\begin{equation}
\label{n_integer}
q_z L = \pi n,
\end{equation}
where $n$ is an integer. It includes the barrier width $L$ and corresponds
to a dimensional phenomenon similar to the Ramsauer-Townsend
effect~\cite{ramsauer1921wirkungsquerschnitt}.
Unlike the classical Ramsauer-Townsend effect, which occurs for a
non-relativistic quantum particle, whose energy exceeds the barrier height,
for Weyl semimetals, below-the-barrier particles may also demonstrate the
same reflectionless transmission. Similar phenomenon has been observed in
graphene: electrons approaching a rectangular barrier at the so-called
`magic angles' propagate through the barrier without
reflection~\cite{chiral_tunn_tudorovskiy2012}.
The Ramsauer-Townsend-like peaks in
$T(k_\bot)$
become more pronounced with the increase of $L$, see
Fig.~\ref{fig::T_vs_L}.

If
relation~(\ref{n_integer})
is violated, coefficient $r$ still can vanish, provided that
\begin{eqnarray}
\label{eq::klein_cond}
k_\bot=\sqrt{m/B}.
\end{eqnarray}
Under this condition
$\varepsilon=k_z=U-q_z$,
the Hamiltonian
\eqref{eq::hamilt} effectively describes a one-dimensional relativistic
particle, and the Klein scattering is observed. Thus, while both 
Eqs.~(\ref{n_integer})
and~(\ref{eq::klein_cond})
correspond to the reflectionless transmission of the incident particle, the
mechanisms responsible for such a transmission are non-identical.

If
$q_z$
is real, the wave function under the barrier is described by a
linear combination of plane waves. From
Eq.~\eqref{eq::qz_def}
we derive that
${\rm Im}\, q_z = 0$
when
\begin{eqnarray}
\frac{m-|\varepsilon-U|}{B}<k_{\bot}^2<\frac{m+|\varepsilon-U|}{B}.
\label{eq::real_qz}
\end{eqnarray}
If this condition is violated, parameter
$q_z$
becomes imaginary, and probability for the electron to pass through the
barrier vanishes exponentially with the growth of $L$. As a result, the
value of $T$ rapidly decays outside the range of
$k_\bot$
defined by
Eq.~(\ref{eq::real_qz}).
The curve at
$\varepsilon/m=0.9$
in
Fig.~\ref{fig::T_vs_k_perp}
illustrates this feature.

In the above discussion we assumed the ballistic regime of the electron
scattering. Such an approach is valid if the electron mean-free path
$l_{\rm mf}$
is larger than the barrier width. For a wider barrier,
$l_{\rm mf}\ll L$,
the electron scattering on the barrier edge at
$z=L$
becomes insignificant, which is equivalent to the limit
$L \rightarrow \infty$,
the case of a
$p$-$n$
junction. Thus, we have no reflected wave within the
barrier and have to match wave function at
$z=0$
only. Solving the system of two linear equations we obtain the expression
for the reflected wave amplitude in the form
\begin{equation}
r
=
\sqrt{\frac{\varepsilon+k_z}{\varepsilon-k_z}}\frac{k_z-U-q_z}{k_z+U+q_z}.
\label{22}
\end{equation}
In the limit
$k_\bot\rightarrow \sqrt{m/B}$,
when
$k_z\rightarrow \varepsilon$,
we obtain
\begin{eqnarray}
r \propto {\sqrt{ \varepsilon - k_z}}.
\end{eqnarray}
Thus, the reflection coefficient vanishes and the Klein tunneling is
observed. Naturally, no `magic angles' exist in the case of the infinite
barrier.

\section{Barrier perpendicular to basal plane}\label{sec::perp}

\begin{figure}[t]
\centering
\includegraphics[width=0.9\linewidth]{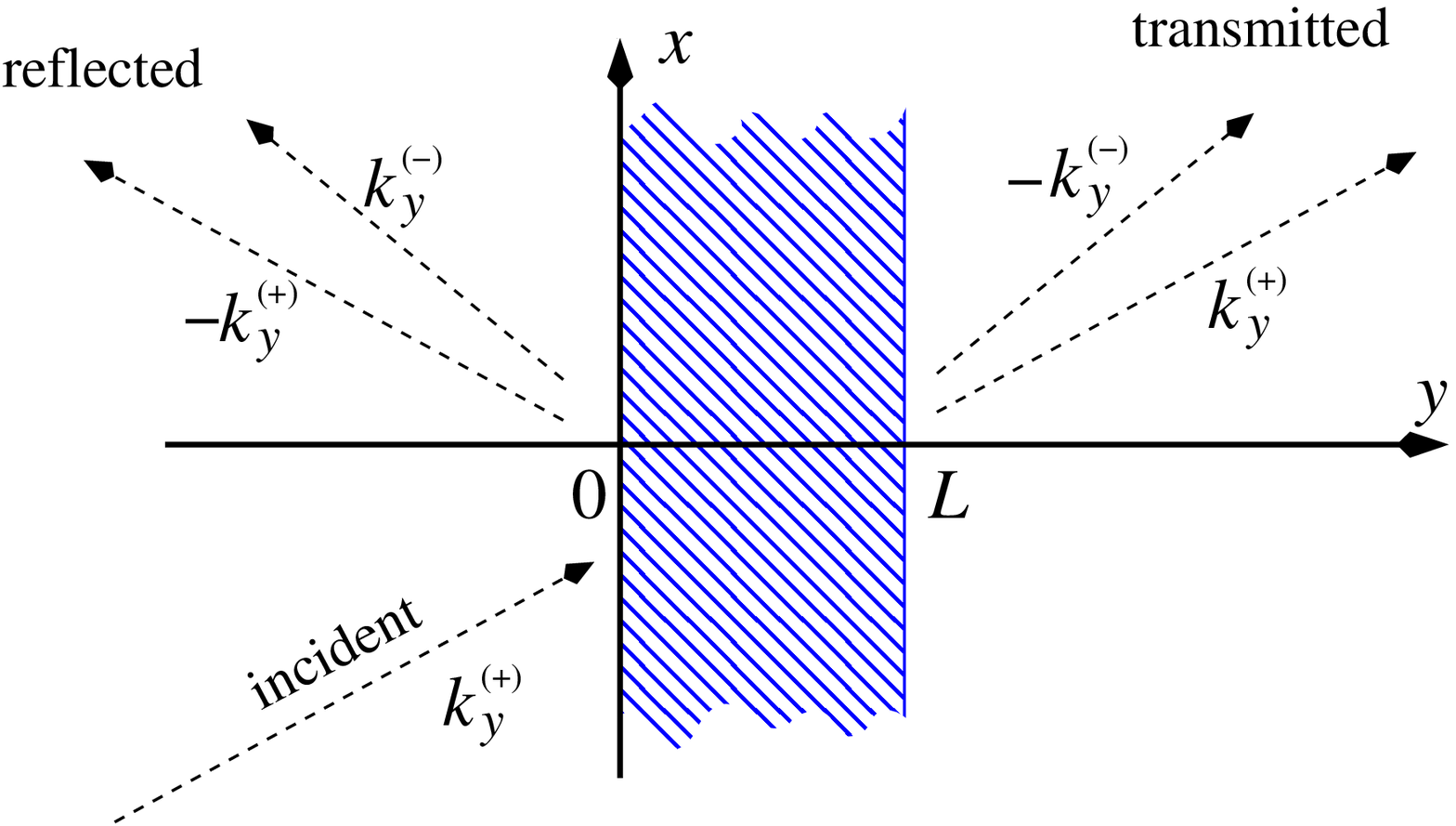}
\includegraphics[width=0.8\linewidth]{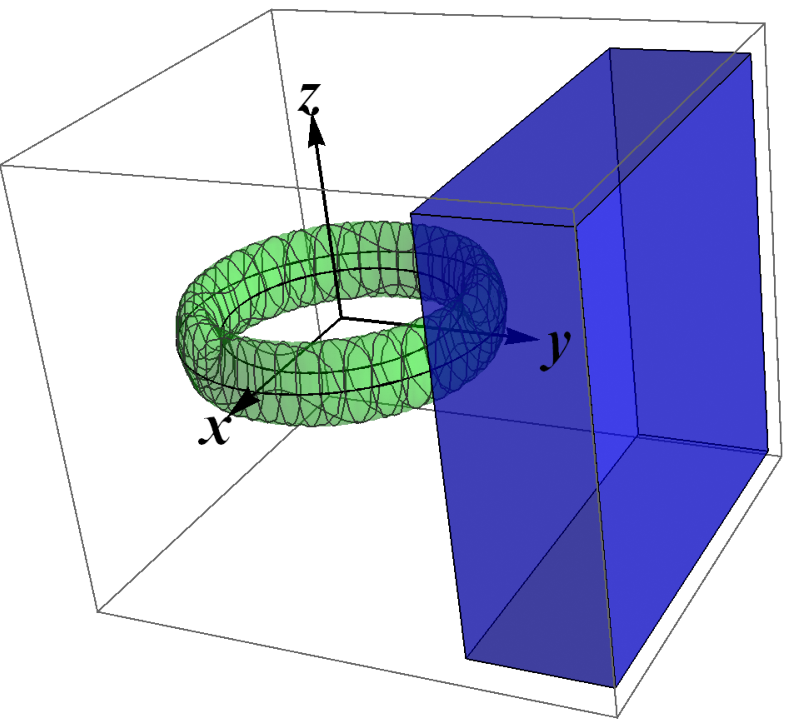}
\caption{Barrier perpendicular to the basal plane. Top panel: Incident,
reflected, and transmitted waves near the barrier. The wave vectors of the
incoming and outgoing waves are shown by the dashed lines with arrows. The
barrier is represented by the blue hatched area. The barrier has finite width
equal to $L$ in the
$y$~direction 
and extends to infinity in the $x$ and $z$ directions.
Bottom~panel: Relative orientation of the barrier and the iso-energy
surface. The iso-energy (green) surface, defined by the equation
$\varepsilon^2 = {(m-Bk_{\bot}^2)^2+k_z^2}$,
is a torus in the reciprocal space. The barrier is shown as blue
rectangular parallelepiped. Within the parallelepiped the potential energy
is $U$, outside it is zero.
\label{Fig3}
}
\end{figure}

Let us consider the situation when the rectangular barrier is perpendicular
to the basal plane. Such a configuration is depicted in
Fig.~\ref{Fig3}.
The nodal ring lies in the
$xy$-plane,
as before. We fix $y$-axis to be normal to the barrier. In such a geometry
momentum components
$k_x$
and
$k_z$
are preserved by the scattering process. As for
$k_y$,
we derive from
Eq.~\eqref{2}
that
$k_y = \pm 
k_{y}^{\scalebox{0.8}{$\scriptscriptstyle (\pm)$}}$,
where
\begin{equation}\label{eq::ky}
k_{y}^{\scalebox{0.8}{$\scriptscriptstyle (\pm)$}} = \sqrt{\frac{1}{B}\left(m-Bk_x^2\pm\sqrt{\varepsilon^2-k_z^2}\right)}.
\end{equation}
We see that four different values of
$k_y$
correspond to a single set of parameters
$(k_x,k_z,\varepsilon)$.
Therefore, an incident electron can be
scattered by the barrier into four possible channels
$k_y = \pm
k_{y}^{\scalebox{0.8}{$\scriptscriptstyle (\pm)$}}$,
see
Fig.~\ref{Fig3}. 
In other words, the flux of incident electrons is distributed between two
transmission channels and two reflection channels.  To distinguish between
the transmission and reflection channels we can use
Eq.~(\ref{eq::current}) 
to make sure that the transmitted particle carries positive current
$j_y$
along $y$-axis. One can easily check that
$k_y =
k_{y}^{\scalebox{0.8}{$\scriptscriptstyle (+)$}}$
and
$k_y =
-k_{y}^{\scalebox{0.8}{$\scriptscriptstyle (-)$}}$
correspond to the transmission channels,
$j_y>0$,
while 
$k_y =
-k_{y}^{\scalebox{0.8}{$\scriptscriptstyle (+)$}}$
and
$k_y =
k_{y}^{\scalebox{0.8}{$\scriptscriptstyle(-)$}}$
correspond to the reflection channel,
$j_y < 0$.
If
$m-Bk_x^2 \geq \sqrt{\varepsilon^2-k_z^2}$
the scattering into four channels is possible. Otherwise, the value of
$k_{y}^{\scalebox{0.8}{$\scriptscriptstyle (-)$}}$
is imaginary, and the transmission and reflection channels corresponding to
$k_{y}^{\scalebox{0.8}{$\scriptscriptstyle (-)$}}$
disappear.

Under the barrier, the wave function is also a linear combination of four
exponentials, each characterized by a specific value of 
$q_y$.
Possible values of
$q_y$
are 
$\pm q_{y}^{\scalebox{0.8}{$\scriptscriptstyle (\pm)$}}$,
where
\begin{equation}
q_{y}^{\scalebox{0.8}{$\scriptscriptstyle (\pm)$}}
=
\sqrt{\frac{1}{B}\left[m-Bk_x^2\pm\sqrt{(\varepsilon-U)^2-k_z^2}\right]}.
\end{equation}

We will confine the following discussion by two constraints. First, we will
assume that the incoming electron is characterized by the momentum
projection
$k_y=k_{y}^{\scalebox{0.8}{$\scriptscriptstyle (+)$}}$.
The incoming electron with
$k_y= - k_{y}^{\scalebox{0.8}{$\scriptscriptstyle (-)$}}$
will not be studied. Second, only the limit
$k_x=0$
will be explicitly discussed. Non-zero
$k_x$
may be easily accounted for by the renormalization of parameter $m$. With
this in mind, we can write the wave function to the left of the barrier
($y<0$)
as a sum of the incident plane wave and two reflected plane waves:
\begin{eqnarray}
\!\!\psi_1 \!=\!e^{i k_{y}^{\scalebox{0.8}{$\scriptscriptstyle (+)$}} y}\!
\begin{pmatrix}
	1\\
	-\chi\\	
\end{pmatrix}\!
+\!
r_{\scalebox{0.8}{$\scriptscriptstyle +$}}
e^{-i k_{y}^{\scalebox{0.8}{$\scriptscriptstyle (+)$}} y} \!
\begin{pmatrix}
	1\\
	-\chi\\	
\end{pmatrix}\!+\!
r_{\scalebox{0.8}{$\scriptscriptstyle -$}}
e^{ik_{y}^{\scalebox{0.8}{$\scriptscriptstyle (-)$}}y}\!
\begin{pmatrix}
	1\\
	\chi\\	
\end{pmatrix}\!,
\label{psi_1_y}
\end{eqnarray}
where
\begin{eqnarray}
\chi=\sqrt{\frac{\varepsilon-k_z}{\varepsilon+k_z}}.
\end{eqnarray}
In the region under the barrier
($0<y<L$),
the wave function can be expressed as
\begin{eqnarray}
\psi_2
&=&
\tilde{a}_{{\scalebox{0.8}{$\scriptscriptstyle -$}}}
e^{iq_{y}^{\scalebox{0.8}{$\scriptscriptstyle (-)$}}y}\!
\begin{pmatrix}
	1\\
	\phi\\
\end{pmatrix}\!
+
\tilde{b}_{\scalebox{0.8}{$\scriptscriptstyle -$}}
e^{-iq_{y}^{\scalebox{0.8}{$\scriptscriptstyle (-)$}}y}\!
\begin{pmatrix}
	1\\
	\phi\\
\end{pmatrix} \!
\\
\nonumber
&+&\tilde{a}_{\scalebox{0.8}{$\scriptscriptstyle +$}}
e^{iq_{y}^{\scalebox{0.8}{$\scriptscriptstyle (+)$}}y}\!
\begin{pmatrix}
	1\\
	-\phi\\
\end{pmatrix}\!
+
\tilde{b}_{\scalebox{0.8}{$\scriptscriptstyle +$}}
e^{-iq_{y}^{\scalebox{0.8}{$\scriptscriptstyle (+)$}}y}
\begin{pmatrix}
	1\\
	-\phi\\
\end{pmatrix},
\label{psi_2_y}
\end{eqnarray}
where
\begin{eqnarray}
\phi=\sqrt{\frac{\varepsilon-U-k_z}{\varepsilon-U+k_z}}.
\end{eqnarray}
Finally, to the right of the barrier
($y>L$),
we have
\begin{eqnarray}
\psi_3\!
=\!
t_{\scalebox{0.9}{$\scriptscriptstyle -$}}
e^{-ik_{y}^{\scalebox{0.8}{$\scriptscriptstyle (-)$}}(y-L)}\!
\begin{pmatrix}
	 1\\
	 \chi\\
\end{pmatrix}\!
+
t_{\scalebox{0.9}{$\scriptscriptstyle +$}}
e^{ik_{y}^{\scalebox{0.8}{$\scriptscriptstyle (+)$}}(y-L)}\!
\begin{pmatrix}
	 1\\
	 -\chi\\
\end{pmatrix}\!.
\label{psi_3_y}
\end{eqnarray}

\begin{figure}[t]
\center{\includegraphics[width=1\linewidth]{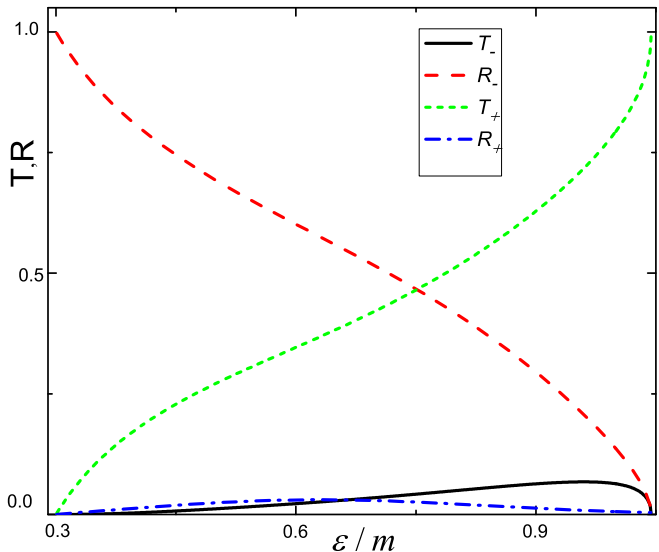}}
\caption{Transmission and reflection coefficients as functions of energy.
The curves are calculated for
$k_z/m=0.3$,
$U/m=1$,
$mL=5$
and
$Bm=1$.
Coefficients
$T_{\pm}(R_{\pm})$
correspond to the transmitted (reflected) waves with
$k_{y}^{(\pm)}$.
We assume that
$k_x=0$
because nonzero
$k_x$
only renormalizes $m$.
\label{Fig4}
}
\end{figure}

\begin{figure}[t]
\center{\includegraphics[width=1\linewidth]{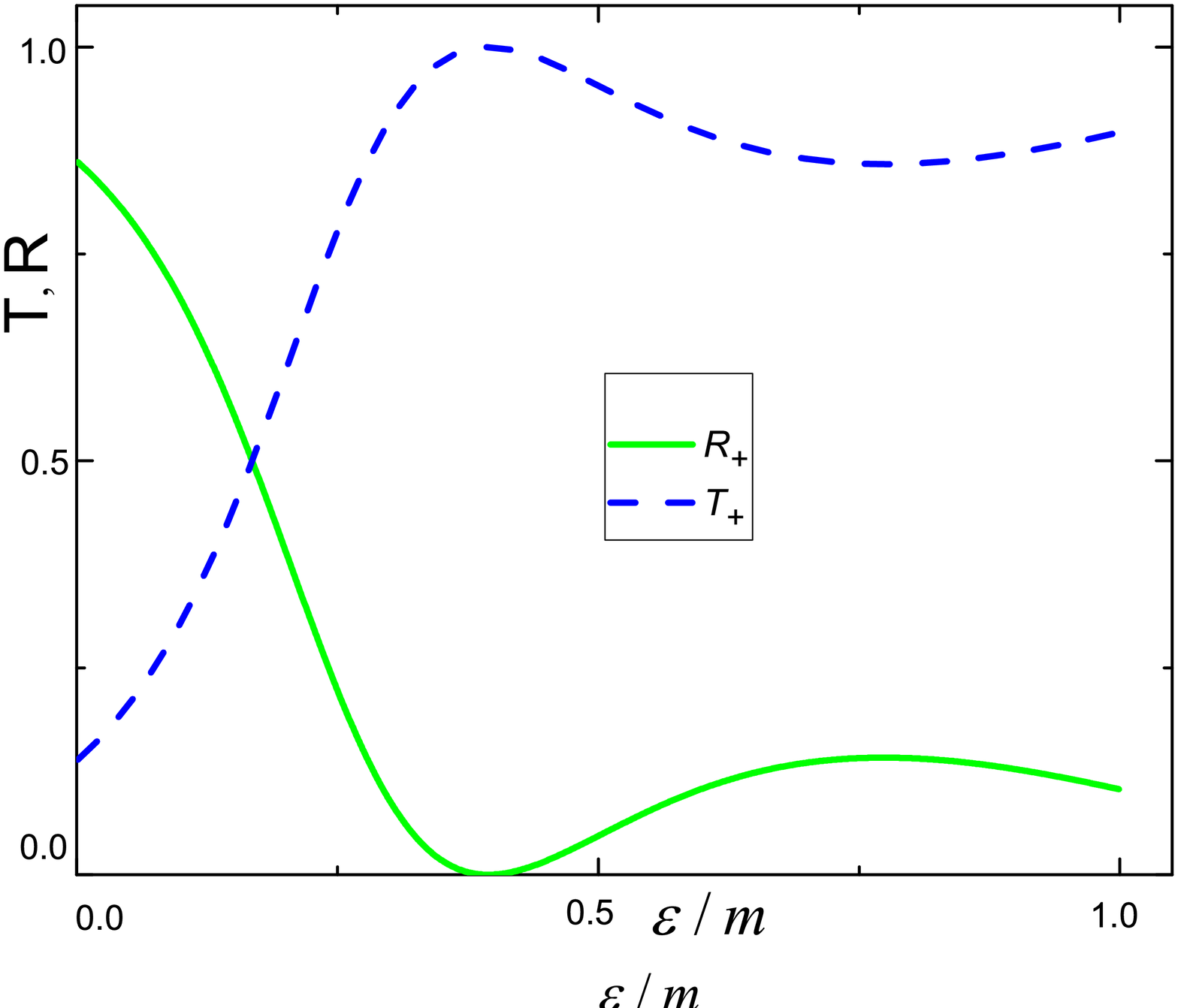}}
\caption{Transmission and reflection coefficients versus energy for the
scattering confined to the basal plane
($k_z = 0$).
The curves are calculated for
$U/m=1$,
$mL=5$
and
$B=1/m$.
Coefficients
$T_{\pm}(R_{\pm})$
describe the transmitted (reflected) waves with
$k_{y}^{(\pm)}$.
In this regime both
$T_-$
and
$R_-$
vanish. We assume that
$k_x=0$
because nonzero
$k_x$
only renormalizes $m$.
\label{Fig5}
}
\end{figure}

The continuity of the probability current at the barrier edges requires the
continuity of the wave function and its $y$-derivative at
$y=0$
and
$y=L$.
As a result, we obtain
\begin{eqnarray}\label{system}
\nonumber
&&(1\!+\!r_{\scalebox{0.8}{$\scriptscriptstyle +$}})\!\!
\begin{pmatrix}
	1\\
	-\chi\\	
\end{pmatrix}
\!+\!
 r_{\scalebox{0.8}{$\scriptscriptstyle -$}}\!
\begin{pmatrix}
	1\\
	\chi\\	
\end{pmatrix}
\!=
(\tilde{a}_{\scalebox{0.9}{$\scriptscriptstyle -$}}\!+\!\tilde{b}_{\scalebox{0.9}{$\scriptscriptstyle -$}})\!\!
\begin{pmatrix}
	1\\
	\phi\\
\end{pmatrix}\!
+
(\tilde{a}_{\scalebox{0.9}{$\scriptscriptstyle +$}}\!+\!\tilde{b}_{\scalebox{0.9}{$\scriptscriptstyle +$}}) \!
\begin{pmatrix}
	1\\
	-\phi\\
\end{pmatrix},
\\
\nonumber
&&k_{y}^{\scalebox{0.8}{$\scriptscriptstyle (+)$}}(1-r_{\scalebox{0.8}{$\scriptscriptstyle +$}}\!)\!
\begin{pmatrix}
	1\\
	-\chi\\	
\end{pmatrix}	
+
r_{\scalebox{0.8}{$\scriptscriptstyle -$}}
k_{y}^{\scalebox{0.8}{$\scriptscriptstyle (-)$}}\!
\begin{pmatrix}
	1\\
	\chi\\	
\end{pmatrix}\!
=\!
q_{y}^{\scalebox{0.8}{$\scriptscriptstyle (-)$}}(\tilde{a}_{\scalebox{0.9}{$\scriptscriptstyle -$}}\!-\!\tilde{b}_{\scalebox{0.9}{$\scriptscriptstyle -$}})
\!
\begin{pmatrix}
	1\\
	\phi\\
\end{pmatrix}
\\
&&+
q_{y}^{\scalebox{0.8}{$\scriptscriptstyle (+)$}}(\tilde{a}_{\scalebox{0.9}{$\scriptscriptstyle +$}}-\tilde{b}_{\scalebox{0.9}{$\scriptscriptstyle +$}})\!\!
\begin{pmatrix}
	1\\
	-\phi\\
\end{pmatrix},
\\
&&\left(\tilde{a}_{\scalebox{0.9}{$\scriptscriptstyle -$}}e^{iq_{y}^{\scalebox{0.8}{$\scriptscriptstyle (-)$}}L}\!+\!\tilde{b}_{\scalebox{0.9}{$\scriptscriptstyle -$}}e^{-iq_{y}^{\scalebox{0.8}{$\scriptscriptstyle (-)$}}L}\right)\!\!
\begin{pmatrix}
	1\\
	\phi\\
\end{pmatrix}
\!+\!
\left(\tilde{a}_{\scalebox{0.9}{$\scriptscriptstyle +$}}e^{iq_{y}^{\scalebox{0.8}{$\scriptscriptstyle (+)$}}L}\!+\!\tilde{b}_{\scalebox{0.9}{$\scriptscriptstyle +$}}
e^{-iq_{y}^{\scalebox{0.8}{$\scriptscriptstyle (+)$}}L}\right)\!\!
\begin{pmatrix}
	1\\
	-\phi\\
\end{pmatrix}
\nonumber
\\
&&=
t_{\scalebox{0.9}{$\scriptscriptstyle -$}}
\begin{pmatrix}
	 1\\
	 \chi\\
\end{pmatrix}
+
t_{\scalebox{0.9}{$\scriptscriptstyle +$}}
\begin{pmatrix}
	 1\\
	 -\chi\\
\end{pmatrix},
\nonumber
\\
&&q_{y}^{\scalebox{0.8}{$\scriptscriptstyle (-)$}}\left(
\tilde{a}_{\scalebox{0.9}{$\scriptscriptstyle -$}}
e^{iq_{y}^{\scalebox{0.8}{$\scriptscriptstyle (-)$}}L}\!-\!\tilde{b}_{\scalebox{0.9}{$\scriptscriptstyle -$}}
e^{-iq_{y}^{\scalebox{0.8}{$\scriptscriptstyle (-)$}}L}
\right)\!\!
\begin{pmatrix}
	1\\
	\phi\\
\end{pmatrix}
\!+\!
\nonumber
\\
&&q_{y}^{\scalebox{0.8}{$\scriptscriptstyle (+)$}}\!\left(
\tilde{a}_{\scalebox{0.9}{$\scriptscriptstyle +$}}
e^{iq_{y}^{\scalebox{0.8}{$\scriptscriptstyle (+)$}}L}\!-\!\tilde{b}_{\scalebox{0.9}{$\scriptscriptstyle +$}}
e^{-iq_{y}^{\scalebox{0.8}{$\scriptscriptstyle (+)$}}L}
\right)\!\!
\begin{pmatrix}
	1\\
\nonumber
	-\phi\\
\end{pmatrix} 	 	
\!=\!
k_{y}^{\scalebox{0.8}{$\scriptscriptstyle (-)$}}
t_{\scalebox{0.9}{$\scriptscriptstyle -$}}\!
\begin{pmatrix}
	 1\\
	 \chi\\
\end{pmatrix}
\!+\!
k_{y}^{\scalebox{0.8}{$\scriptscriptstyle (+)$}}
t_{\scalebox{0.9}{$\scriptscriptstyle +$}}\!
\begin{pmatrix}
	 1\\
	 -\chi\\
\end{pmatrix}\!.
\end{eqnarray}
Solving these equations numerically, we calculate the amplitudes
$r_{\scalebox{0.8}{$\scriptscriptstyle \pm$}}$,
$t_{\scalebox{0.8}{$\scriptscriptstyle \pm$}}$,
and obtain two transmission coefficients
$T_{\scalebox{0.9}{$\scriptscriptstyle \pm$}}
=
|t_{\scalebox{0.9}{$\scriptscriptstyle \pm$}}|^2$
and two reflection coefficients
$R_{\scalebox{0.9}{$\scriptscriptstyle \pm$}}
=
|r_{\scalebox{0.9}{$\scriptscriptstyle \pm$}}|^2$.
Particle conservation implies that
$T_{\scalebox{0.9}{$\scriptscriptstyle +$}}
+
T_{\scalebox{0.9}{$\scriptscriptstyle -$}}
+
R_{\scalebox{0.9}{$\scriptscriptstyle +$}}
+
R_{\scalebox{0.9}{$\scriptscriptstyle -$}} = 1$.
The results of the calculations are shown in
Figs.~\ref{Fig4}
and~\ref{Fig5},
where transmission and reflection coefficients are plotted as functions of
$\varepsilon$ for different parameter values.

When the energy $\varepsilon$ is within the interval
$k_z<\varepsilon<\sqrt{k_z^2+m^2}$,
the quantities
$k_{y}^{\scalebox{0.8}{$\scriptscriptstyle (\pm)$}}$
are all real. Consequently, all four scattering channels are open. Under
this conditions the scattering of the incident wave with
$k_{y}^{\scalebox{0.8}{$\scriptscriptstyle (+)$}}$
to the reflected wave with
$k_{y}^{\scalebox{0.8}{$\scriptscriptstyle (-)$}}$
may be significant. It becomes particularly strong if $\varepsilon$ is
close to
$k_z$:
in this regime
$R_- \rightarrow 1$
when
$\varepsilon \rightarrow k_z$,
see
Fig.~\ref{Fig4}.
At the opposite end of the considered energy interval,
$\varepsilon \rightarrow \sqrt{k_z^2+m^2}$,
the reflection probabilities vanish, and the incident particle passes
through the barrier without reflection, preserving its momentum. 

An analytical expressions for the transmission and reflection coefficients
can be obtained in the limit
$k_z=0$,
when the incident particle momentum lies in $xy$-plane. (Since 
$k_z$
is conserved, the momenta of the transmitted and reflected particles are
confined to the basal plane as well.) In such a situation 
Hamiltonian~(\ref{eq::hamilt}) 
decouples into two copies of a scalar non-relativistic Hamiltonian with the
spectrum
${\varepsilon(k_y,k_z=0)=\pm (m-Bk_y^2)}$,
and
$\chi=\phi=1$.
It is easy to check that a particle with the momentum component
$k_{y}^{\scalebox{0.8}{$\scriptscriptstyle (+)$}}
=
\sqrt{\left(m+\varepsilon\right)/B}$
cannot be scattered to
$k_{y}^{\scalebox{0.8}{$\scriptscriptstyle (-)$}}$
channel since
$k_{y}^{\scalebox{0.8}{$\scriptscriptstyle (+)$}}$
and
$k_{y}^{\scalebox{0.8}{$\scriptscriptstyle (-)$}}$
belong to different sectors. Solving
Eqs.~\eqref{system} 
in this limit, we obtain
$R_{\scalebox{0.8}{$\scriptscriptstyle -$}}
=T_{\scalebox{0.8}{$\scriptscriptstyle -$}} = 0$
and
\begin{eqnarray}
\label{eq::non_rel}
R_{\scalebox{0.8}{$\scriptscriptstyle +$}}
=
\frac{
	[(q_{y}^{\scalebox{0.8}{$\scriptscriptstyle (-)$}})^2
	-
	(k_{y}^{\scalebox{0.8}{$\scriptscriptstyle (+)$}})^2]
	\sin^2(q_y^{\scalebox{0.8}{$\scriptscriptstyle (-)$}}L)
     }
     {
	4(q_{y}^{\scalebox{0.8}{$\scriptscriptstyle (-)$}}
	  k_{y}^{\scalebox{0.8}{$\scriptscriptstyle (+)$}})^2
	+
	[(q_{y}^{\scalebox{0.8}{$\scriptscriptstyle (-)$}})^2
	-
	(k_{y}^{\scalebox{0.8}{$\scriptscriptstyle (+)$}})^2]
	\sin^2(q_{y}^{\scalebox{0.8}{$\scriptscriptstyle (-)$}}L)
     },
\end{eqnarray}
where
$q_{y}^{\scalebox{0.8}{$\scriptscriptstyle (-)$}}
=
\sqrt{\left(m-|\varepsilon-U|\right)/B}$.
The dependencies of
$R_{\scalebox{0.8}{$\scriptscriptstyle +$}}$
and
$T_{\scalebox{0.8}{$\scriptscriptstyle +$}}
=
1-R_{\scalebox{0.8}{$\scriptscriptstyle +$}}$
versus
$\varepsilon$
are shown in
Fig.~\ref{Fig5}.
These functions are non-monotone due to dimensional oscillating factor
$\sin^2(q_y^{\scalebox{0.8}{$\scriptscriptstyle (-)$}}L)$.
Equation~(\ref{eq::non_rel})
is, in some respects, similar to the expression describing scattering of a
non-relativistic particle on a rectangular
barrier~\cite{Griffiths}.
However, there is an important difference. The transmission coefficient of
a non-relativistic particle
$T_{\rm nrel} (\varepsilon)$
oscillates due to dimensional effect if
$\varepsilon > U$,
but is monotone if
$\varepsilon < U$.
In the case of the nodal-ring semimetal, the functions
$R_{\scalebox{0.8}{$\scriptscriptstyle +$}}(\varepsilon)$
and, consequently,
$T_{\scalebox{0.8}{$\scriptscriptstyle +$}}(\varepsilon)$
are non-monotone even for 
$\varepsilon < U$.
Mathematically, this occurs due to the existence of the plane wave
solutions with
${\rm Im}\, q_{y}^{\scalebox{0.8}{$\scriptscriptstyle (-)$}}
=0$
in the regime
$\varepsilon<U$.

The case of the infinite-length barrier can be considered in the same
manner. After matching the wave function and its derivative at the barrier
edge, reflection coefficients
$R_{\scalebox{0.8}{$\scriptscriptstyle \pm$}}$
can be calculated. In general, both of them are non-zero, that is, the
scattering in four channels is possible. In the limit
$k_z = 0$
we can find explicit formulas
\begin{eqnarray}
R_{+}&\!=\!&\left(\!\frac{k_y^{(+)}\!-\!q_y^{(+)}}{k_y^{(+)}\!+\!q_y^{(+)}}\!\right)^2\!,\,\,\, T_{+}\!=\!\left(\!\frac{2k_y^{(+)}}{k_y^{(+)}\!+\!q_y^{(+)}}\!\right)^2\!,\\ \nonumber
R_{-}&\!=\!&T_{-}=0.
\label{eq::non_rel_inf}
\end{eqnarray}
This result is similar to the case of the scattering of non-relativistic particle.

\section{Conclusion}\label{sec::discussion}

We show that the electron scattering in the nodal-line semimetals
demonstrates unusual features, such as the Klein tunneling, reflectionless
transmission at `magic angles' (which is an analogue of the classical
Ramsauer-Townsend effect), and the emergence of the additional scattering
channels.

The Klein tunneling occurs for a barrier parallel to the basal plane. The
momentum of the incident particle must satisfy the condition
$m-Bk_{\bot}^2=0$.
If these requirements are met, 
Hamiltonian~\eqref{eq::hamilt} 
effectively describes one-dimensional massless fermions, for which the
Klein tunneling is a well-established phenomenon.

Besides the Klein tunneling, reflectionless propagation across the barrier
can be observed for a particle colliding with the barrier at certain `magic
angles'. These angles depend on the barrier width. Such a behavior is
related to the Ramsauer-Townsend effect for a non-relativistic quantum
particle. Similar phenomena discussed for 
graphene~\cite{chiral_tunn_tudorovskiy2012}.
However, the classical Ramsauer-Townsend effect exists for a particle
whose energy exceeds the height of the barrier, while in the nodal-line
semimetals a particle with
$\varepsilon < U$
also demonstrates the same reflectionless propagation. 

When the barrier is perpendicular to the basal plane, the Klein tunneling
is impossible. In this configuration another interesting phenomenon can be
observed: the second scattering channel becomes available both for
transmitted and reflected particles. Such an unusual scattering occurs
because the system of equations describing conservation of the electron
energy and momentum has two different roots. Therefore, two different
values of 
$|k_y|$
are admissible. The first of these values is the same as 
$|k_y|$ 
of the incident particle, while the second differs. Thus, the wave
functions of the transmitted and reflected particles are superpositions of
two states with unequal momenta. Depending on the scattering parameters,
the probability of changing 
$|k_y|$
after a scattering event can be substantial. We prove this for scattering
processes confined to the basal plane. 

\section*{Acknowledgments}  This work was supported by the Presidium of RAS
(Program I.7, Modern problems of photonics, the probing of inhomogeneous
media and materials).


\end{document}